\documentclass[final,3p,times,twocolumn,authoryear]{elsarticle}

\usepackage{amssymb}
\usepackage{amsmath}

\usepackage{xcolor}
\definecolor{light-gray}{gray}{0.95}

\usepackage{multirow}
\usepackage{arydshln}
\usepackage{url}
\usepackage{caption}
\usepackage{subcaption}
\usepackage{listings} 

\usepackage{hyperref}       
\usepackage[capitalise,noabbrev]{cleveref}  
\lstdefinestyle{yaml}{
     basicstyle=\color{blue}\ttfamily\footnotesize,
     rulecolor=\color{black},
     numbers=left,
     numberstyle=\color{black},
     string=[s]{'}{'},
     stringstyle=\color{blue},
     comment=[l]{:},
     commentstyle=\color{black},
     morecomment=[l]{-},
     breaklines=true,
     backgroundcolor = \color{light-gray}
 }

\journal{Computer Speech and Languages}

\makeatletter
\def\ps@pprintTitle{%
  \let\@oddhead\@empty
  \let\@evenhead\@empty
  \def\@oddfoot{\hfill Accepted for publication in \textit{Computer Speech \& Language}, 
    Vol.~96 (2026), Article~101869, ISSN~0885-2308.\hfill}
  \let\@evenfoot\@oddfoot}
\makeatother

\begin{document}

\begin{frontmatter}



\title{Comparative Study on Noise-Augmented Training and its Effect on Adversarial Robustness in ASR Systems} 


\author[label1,label2]{Karla Pizzi\corref{contrib}} 
\author[label3]{Mat\'ias~Pizarro\corref{contrib}} 
\author[label3]{Asja Fischer} 

\cortext[contrib]{Authors contributed equally.}

\affiliation[label1]{organization={Neodyme AG},
            city={Garching},
            country={Germany}}
\affiliation[label2]{organization={Technical University Munich}, 
            city={Garching},
            country={Germany}}
\affiliation[label3]{organization={Faculty of Computer Science, Ruhr University Bochum},
            city={Bochum},
            country={Germany}}

\begin{abstract}
    In this study, we investigate whether noise-augmented training can concurrently improve adversarial robustness in automatic speech recognition (ASR) systems. 
    We conduct a comparative analysis of the adversarial robustness of four different ASR architectures, each trained under three different augmentation conditions: (1) background noise, speed variations, and reverberations; (2) speed variations only; (3) no data augmentation.
    We then evaluate the robustness of all resulting models against attacks with white-box or black-box adversarial examples.
    Our results demonstrate that noise augmentation not only enhances model performance on noisy speech but also improves the model's robustness to adversarial attacks. 
\end{abstract}


\end{frontmatter}



\section{Introduction}\label{sec1}
Adversarial attacks, also known as evasion attacks, on machine learning models have gathered a lot of attention due to their potential to compromise the models' reliability and security~\citep{szegedy2013intriguing}. 
Initially, research primarily focused on adversarial attacks targeting computer vision models, but there has been a growing interest in the vulnerability of audio processing models, especially speech processing systems, to such attacks.
At the same time, efforts to enhance noise robustness have been made to improve the usability of speech processing systems in various environments~\citep{dua2023noise}. 
However, the relationship between noise robustness and adversarial robustness in the audio domain remains largely unexplored. 

This paper builds upon and extends our previous work~\citep{pizzi2024reassessing} by offering a broader and more rigorous evaluation of the relationship between noise robustness and adversarial robustness in ASR systems. 
Beyond the earlier study, which focused solely on white-box attacks, the current work incorporates both white-box and black-box adversarial attacks and includes the SI-SDR metric for a perceptual assessment of distortion. 
These additions enable a more comprehensive and generalizable analysis of how noise-augmented training affects adversarial robustness across diverse threat models and model types.

Noise robustness, also known as environmental robustness~\cite{ko2017study}, reflects a model's ability to maintain performance despite exposure to various forms of noise and perturbations. 
Given that real-world audio data often includes inherent noise and environmental disturbances, ensuring noise robustness is essential for reliable audio processing and analysis~\citep{virtanen2012techniques}.

Adversarial robustness, on the other hand, refers to a model's ability to resist adversarial attacks and maintain accurate predictions when faced with manipulated audio samples. 
These attacks introduce perturbations that are often imperceptible to humans but can deceive the model into making incorrect predictions~\citep{carlini2018audio, qin2019imperceptible}. 
In automatic speech recognition (ASR) systems, adversarial attacks can compromise security and privacy, potentially leading to unauthorized access or incorrect processing of sensitive data. 
For instance, an attacker could alter an audio signal to mislead an ASR system into executing commands, such as purchasing an unwanted product or navigating to a malicious webpage~\citep{schonherr2019adversarial}. 
Thus, adversarial robustness is critical for model security, enabling the detection and prevention of malicious attacks on audio systems~\citep{olivier2023assessing}.

Understanding the interplay between noise robustness and adversarial robustness in the audio domain is crucial~\citep{olivier2022there}. 
We hypothesize that ASR models with enhanced resistance to acoustic perturbations may inherently exhibit a degree of robustness to adversarial attacks. 
To test this hypothesis, we conduct a comprehensive empirical analysis, evaluating the impact of noise-augmented training on the adversarial defense capabilities of various high-performance ASR models. 
Our study encompasses a range of noise types, such as reverberations and background disturbances, and assesses the models' robustness against different forms of adversarial evasion attacks, including both white-box and black-box scenarios.

\section{Background and Related Work}\label{sec2}
This section provides an overview of various types of adversarial attacks against ASR systems as well as of previous results on noise and adversarial robustness of ASR systems.
Additionally, we discuss what is known about the interdependence between adversarial and noise robustness, drawing on literature from the image domain.

\subsection{Adversarial Attacks}\label{subsec1}
Adversarial examples are inputs intentionally altered to cause machine learning models to make errors. 
These perturbations are usually imperceptible to humans and can significantly degrade model performance. 
In the context of speech recognition, adversarial examples exploit the vulnerabilities of ASR systems, resulting in incorrect transcriptions. 
Two different classes of adversarial attacks exist, i.e., targeted and untargeted attacks, as described in the following. 

\subsubsection{Targeted Attacks}\label{subsubsec1} 
\begin{figure*}[!t]
    \centering
    \begin{subfigure}[b]{0.32\textwidth}
        \centering
        \includegraphics[width=\textwidth]{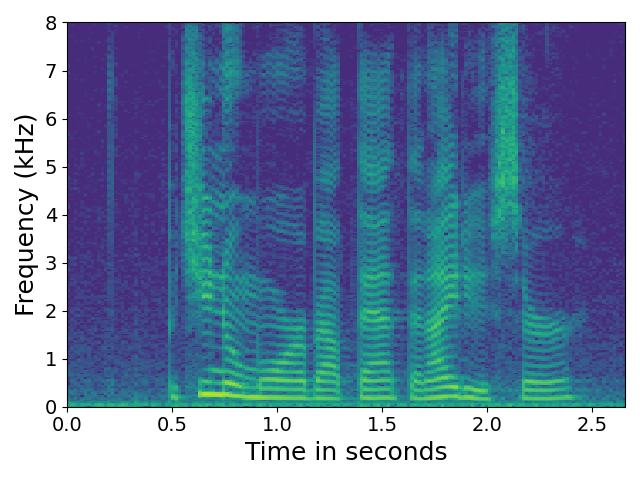}
        \caption{Original audio signal. Transcription:
        ``but you know more about that than i do sir''.}
    \end{subfigure}
    \hfill
    \begin{subfigure}[b]{0.32\textwidth}
        \centering
        \includegraphics[width=\textwidth]{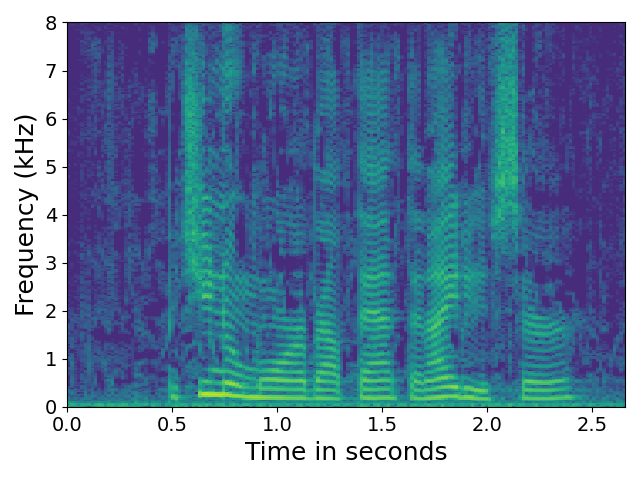}
        \caption{C\&W adversarial example. Transcription:
        ``yes my dear watson i have solved the mystery''.}
    \end{subfigure}
    \hfill
    \begin{subfigure}[b]{0.32\textwidth}
        \centering
        \includegraphics[width=\textwidth]{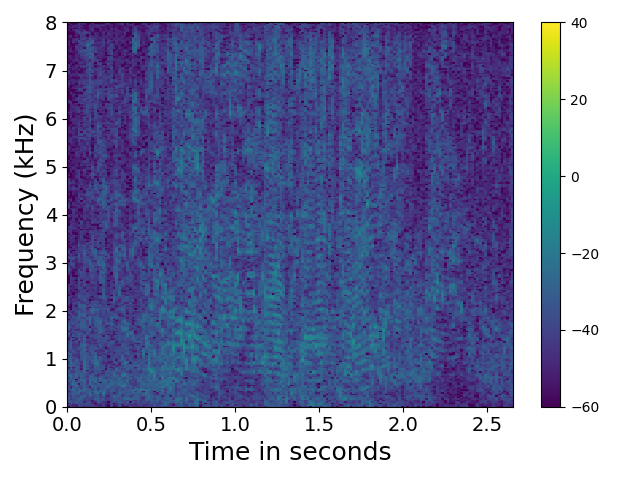}
        \caption{Adversarial noise signal $\delta$. Difference between (a) and (b).}
    \end{subfigure}
    \caption{The spectrogram of (a) the original audio signal is compared to (b) the spectrogram of its corresponding C\&W adversarial example, and (c) the spectrogram of the adversarial noise.
    }
    \label{fig:c_w}
\end{figure*}

Targeted attacks on ARS systems aim at creating manipulated audio data that is transcribed to a specific target phrase chosen by the attacker. These attacks are designed for a white-box setting, which means that the attacker must have full access to the model. 
\citet{carlini2018audio} introduced the first efficient white-box targeted attack against ASR systems, known as the C\&W~attack.
Originally designed for Mozilla's DeepSpeech model~\citep{Hannun/2014/deepspeech}, the C\&W~attack computes the adversarial perturbation $\delta$ for a given benign audio sample $x$ by performing gradient descent on the following loss function: 
\begin{equation*}
  L(x, \delta, \hat y) = l_{m}(f(x+\delta), \hat y) + c \cdot |\delta|_2^2 \enspace,
\end{equation*}
where $f(\cdot)$ is the function represented by the ASR model mapping the audio input to the most likely transcription, $l_{m}$ is the loss function measuring the accuracy of the ASR system's prediction w.r.t.\ the target transcription $\hat y$, and $c$ is a hyperparameter regulating the strength of penalization for a larger perturbation. 
During gradient-based minimization, the perturbation $\delta$ (also referred to as adversarial noise) is constrained to keep smaller than a predefined value $\epsilon$. 
This $\epsilon$ value is decreased iteratively to minimize the distortion. 
Moreover, the hyperparameter $c$ is initially set to a small value to simplify finding an adversarial example first. 
It is then increased to minimize the needed perturbance. 
The resulting adversarial audio files closely resemble the original samples in terms of their dB scale.
However, they lead to a different transcription output by the ASR system.
Figure~\ref{fig:c_w} compares the original spectrum of a benign audio signal with its corresponding C\&W adversarial counterpart.

\subsubsection{Untargeted Attacks}\label{subsubsec2}
Untargeted attacks aim to cause a machine learning model to produce an incorrect output without specifying what the incorrect output should be. 
Unlike targeted attacks, where the adversary has a specific erroneous outcome in mind (e.g., transcribing an audio sample to a specific phrase), untargeted attacks are concerned only with causing the model to fail in its prediction.
Untargeted attacks can be particularly useful in scenarios where the goal is to broadly disrupt the functionality of a system or to demonstrate the general vulnerability of a model to adversarial manipulation. 
These attacks do not require the attacker to have detailed knowledge of the model's architecture or parameters. 
In this work, we consider two untargeted black-box attacks for ASR systems~\citep{alzantot2018did, abdullah2021hear}.
These attacks rely on probing the model with different inputs to generate adversarial examples that lead to incorrect transcriptions without needing access to the internal workings of the model, which makes them particularly interesting under certain attack scenarios. 
We refer to these attacks as Alzantot and Kenansville attack, respectively.

\begin{figure*}[t!]
    \centering
    \begin{subfigure}[b]{0.32\textwidth}
        \centering
        \includegraphics[width=\textwidth]{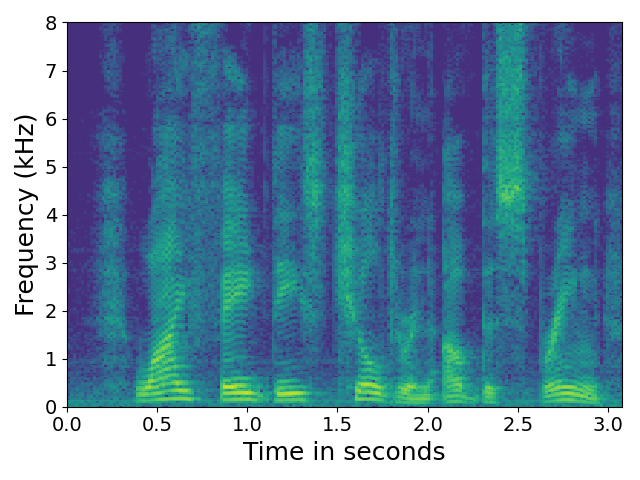}
        \caption{Original audio signal. Transcription:
        ``why fade these children of the spring''.}
    \end{subfigure}
    \hfill
    \begin{subfigure}[b]{0.32\textwidth}
        \centering
        \includegraphics[width=\textwidth]{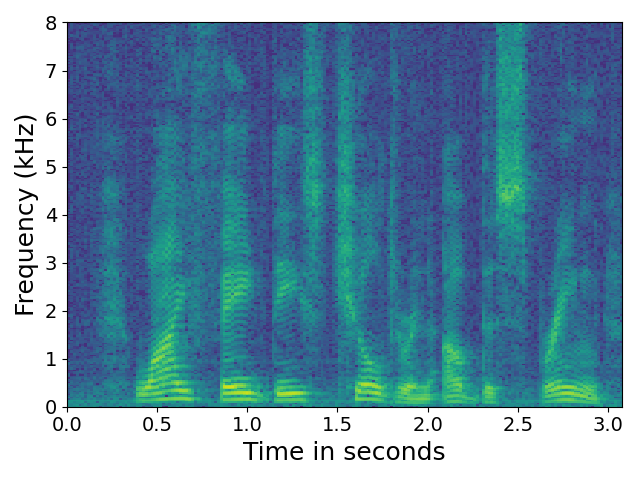}
        \caption{Alzantot adversarial example. Transcription:
        ``why fague these children of the spring''.}
    \end{subfigure}
    \hfill
    \begin{subfigure}[b]{0.32\textwidth}
        \centering
        \includegraphics[width=\textwidth]{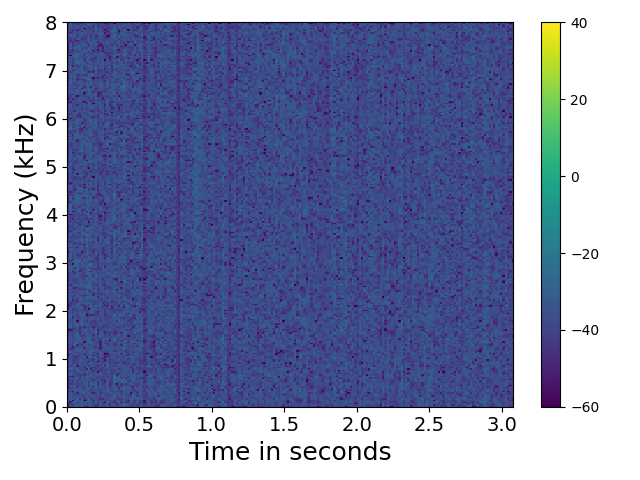}
        \caption{Adversarial noise signal $\delta$. Difference between (a) and (b).}
    \end{subfigure}
        \caption{The spectrogram of (a) the original audio signal is compared to (b) the spectrogram of its corresponding Alzantot adversarial example, and (c) the spectrogram of the adversarial noise.
        }
        \label{fig:alzantot}
\end{figure*}

\begin{figure*}[t!]
    \centering
    \begin{subfigure}[b]{0.32\textwidth}
        \centering
        \includegraphics[width=\textwidth]{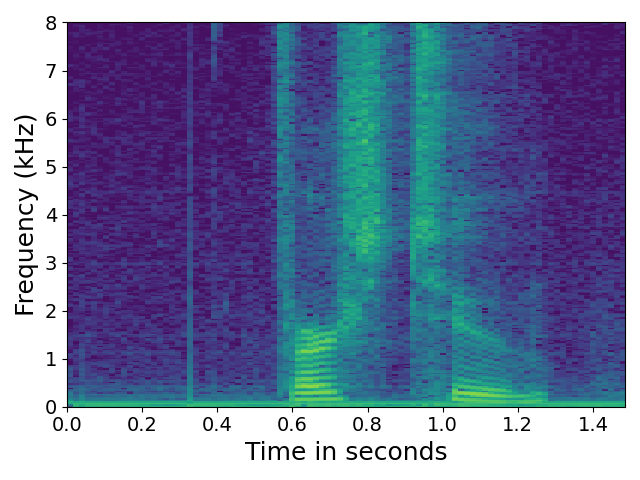}
        \caption{Original audio signal.  Transcription: `` verse two ''.}
    \end{subfigure}
    \hfill
    \begin{subfigure}[b]{0.32\textwidth}
        \centering
        \includegraphics[width=\textwidth]{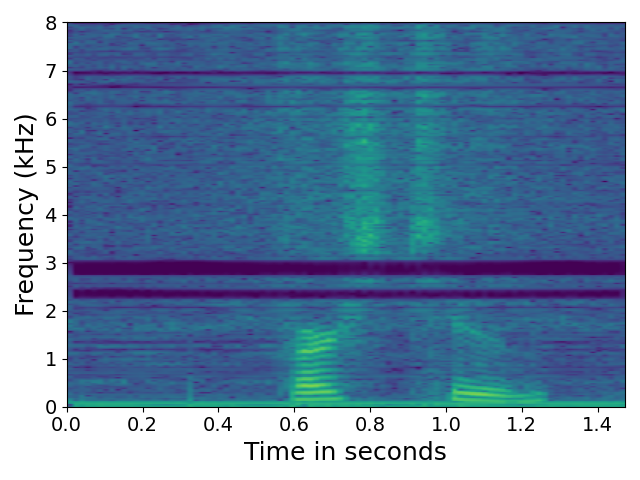}
        \caption{Kenansville adversarial example. Transcription:
        ``first too''.}
    \end{subfigure}
    \hfill
    \begin{subfigure}[b]{0.32\textwidth}
        \centering
        \includegraphics[width=\textwidth]{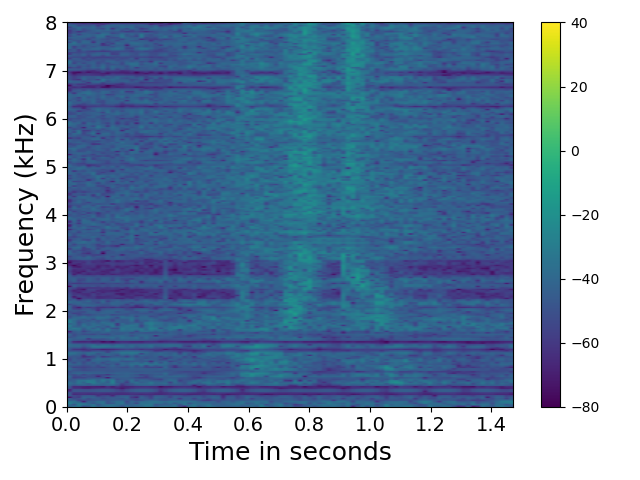}
        \caption{Adversarial noise signal $\delta$. Difference between (a) and (b).}
    \end{subfigure}
        \caption{The spectrogram of (a) the original audio signal is compared to (b) the spectrogram of its corresponding Kenansville adversarial example, and (c) the spectrogram of the adversarial noise.
        }
        \label{fig:kenansville}
\end{figure*}

While \cite{alzantot2018did} use genetic algorithms, an optimization technique that simulates the process of natural selection by repeatedly selecting, combining, and mutating candidate solutions to find the most optimal or fit solution to a problem, \cite{abdullah2021hear} present an attack leveraging signal processing methods by systematically removing low-intensity frequency components at each step.
They try to remove a minimal number of frequencies that would cause a mismatch between the model's output and the original label.
While the Kenansville attack is actually black-box -- they just use the model's output --, the Alzantot attack can be enhanced with access to the loss function to improve the success rate, which would make it more of a gray-box approach~\citep{olivier2022recent}. 
Figures~\ref{fig:alzantot}~and~\ref{fig:kenansville} illustrate a comparison between the original spectrum of a benign audio signal and the corresponding adversarial counterparts created with the Alzantot and Kenansville attack, respectively.

\subsection{Model Robustness}
This paper considers two notions of robustness: noise robustness and adversarial robustness. 
We evaluate how noise robustness, achieved through noise-augmented training, impacts adversarial robustness.

\subsubsection{Noise Robustness}
Environmental noise is a common challenge in real-world audio processing, requiring ASR systems to handle disturbances like background chatter, traffic noise, and reverberation. 
General noise, similar to adversarial noise, can cause non-targeted misclassifications, affecting accuracy and user experience. 
Strategies to enhance noise robustness include noise augmentation, denoising algorithms, and robust feature extraction techniques~\citep{abe2015spectralSubtraction, stoller2018-waveunet, bagheri219WienerFilter, alamdari2021improving}.

Noise-augmented training is widely recognized as a reliable method for improving noise resilience~\citep{ko2017study}. 
This technique integrates noise-perturbed data during neural network training by adding specific noise types to clean data. 
Modern ASR systems, such as SpeechBrain~\citep{ravanelli2021speechbrain}, use noise augmentation in their training protocols.
The selection and weighting of noise types are crucial for the method's effectiveness.

While some approaches focus on architectural modifications to inherently enhance noise robustness, such as changes in hidden layers or input data representation~\citep{dua2023noise}, we focus on noise-augmented training without altering model architecture or processing pipelines. 

\subsubsection{Adversarial Robustness}
Adversarial robustness concerns a model's ability to resist adversarial attacks and maintain accurate predictions when faced with manipulated audio samples, as outlined in Section~\ref{subsec1}. 
Various techniques have been proposed to enhance this robustness, including adversarial training, defensive distillation, gradient masking, regularization techniques, robust optimization, and feature denoising during training~\citep{goodfellow2014explaining, madry2018towards, papernot2016distillation, tramer2018ensemble, jakubovitz2018improving, wong2018provable, xie2019feature}. 
During inference, techniques like input transformations and randomization may be employed~\citep{xu2018feature, pizarro2021robustifying}.

Adversarial training involves training the model on both clean and adversarially perturbed samples to enhance its robustness against adversarial manipulations. 
However, this process is considered computationally intensive, especially for large datasets. 
\citet{zhang2019limitations} highlight that while adversarial training is effective for simple datasets, it struggles to scale for high-dimensional data like speech, leaving many blind spots.

In this work, we evaluate whether noise robustness can enhance adversarial robustness. 
If successful, this approach could potentially reduce the computational resources needed for generating adversarial examples during training, thereby minimizing blind spots without compromising robustness.

\subsubsection{Correlation between Robustness Notions}
Noise robustness enables a model to maintain accuracy amid naturally occurring acoustic disturbances, whereas adversarial robustness refers to robustness against maliciously engineered inputs. 
The relationship between these robustness types in the image domain has been debated: some studies argue for their independence~\citep{fawzi2016robustness, laugros2019adversarial}, while others suggest a synergistic relationship~\citep{gilmer2019adversarial, li2019certified}. 
The consensus on their interrelation remains elusive.

Furthermore, the conclusions from image-based research do not directly apply to the audio domain~\citep{zelasko2021adversarial}. 
ASR systems possess unique characteristics, and audio adversarial attacks can exploit phenomena like psychoacoustic hiding, absent in visual contexts~\citep{schonherr2019adversarial}. 
While many studies focus on one type of robustness~\citep{ko2017study, qin2019imperceptible, schonherr2019adversarial}, few have examined both in the audio context~\citep{olivier2022recent, Pizarro2024DistriBlock}.

Driven by the ongoing debate in the image domain and the lack of comprehensive studies in the audio domain, this research investigates the interrelation between adversarial and noise robustness in ASR models. 
We evaluate features related to the success of adversarial example generation and the effort required to produce these examples.

\section{Methodology}
In this section, we describe the methodology used in our experiments.

\subsection{ASR Models}
We analyze end-to-end ASR systems, which directly transform spoken language into text using a single, unified model, thus differing from traditional systems with separate stages~\citep{E2E_Survey/2024/IEEE/ACM}. 
We utilize SpeechBrain, an open-source platform for speech technologies~\citep{ravanelli2021speechbrain}. 
SpeechBrain ASR systems typically consist of three components: (i) during training, a \textbf{tokenizer} breaks the continuous stream of text into discrete normalized units, called tokens (usually characters or subwords) and during inference maps output tokens back to text; (ii) the \textbf{acoustic model} maps feature vectors derived from the raw input speech into probability distributions over the tokens; and (iii) an optional \textbf{language model} that enhances prediction accuracy by evaluating the most likely sequences.

We analyse four different architectures from the SpeechBrain toolkit (using the categories from the SpeechBrain recipes): 
\begin{itemize}
    \item \textbf{CTC:} Uses a unigram tokenizer encoding 31 characters representing the English alphabet and the well-established pre-trained transformer-based model \emph{wav2vec 2.0}\footnote{For the wav2vec~2.0 pre-trained model, see \url{https://huggingface.co/facebook/wav2vec2-large-960h-lv60-self}.} as the acoustic model~\citep{baevski2020wav2vec}, a well-established framework for self-supervised learning of speech representations to extract audio features.
    This model is subsequently fine-tuned by integrating it with a vanilla deep neural network (DNN) that uses a softmax function to learn discrete speech units corresponding to the tokens. 
    The DNN is trained with Connectionist Temporal Classification (CTC) loss~\citep{Graves/2006/ICML}. 
    Since the alignment between the input speech frames with their corresponding textual output is based on CTC, it does not require pre-segmented data, i.e., data divided into smaller chunks. 
    This approach enables the model to learn the alignment explicitly, facilitating the training process for ASR tasks.
    \item \textbf{seq2seq~1:} Uses a unigram tokenizer, which transforms words into subwords represented in 5000 units.
    Additionally, it employs an encoder, a decoder, and an attention mechanism between them as the acoustic model. 
    The encoder consists of a combination of convolutional, recurrent, and fully-connected networks, commonly known as CRDNN in the literature~\citep{ravanelli2021speechbrain}.
    The decoder employs a long short-term memory network~\citep{Hochreiter/1997/MIT} coupled with an attention mechanism~\citep{Chorowski/2015/NIPS}, initially introduced in the Listen, Attend, and Spell system~\citep{William/2016/ICASSP}.
    It incorporates a pre-trained language model, trained on a transformer model\footnote{For the transformer pre-trained language model, see \url{https://speechbrain.readthedocs.io/en/latest/API/speechbrain.lobes.models.transformer.TransformerLM.html}.} using the entire LibriSpeech corpus containing approximately 10 million words. 
    This architecture is particularly suited for tasks where the input and output sequences are of variable lengths, making it a popular choice for ASR systems.
    \item \textbf{seq2seq~2:} Similar to seq2seq~1, but with a tokenizer with only 1000 subword units and a pre-trained recurrent neural network-based language model  \footnote{For the RNN pre-trained language model, see \url{https://speechbrain.readthedocs.io/en/latest/API/speechbrain.lobes.models.RNNLM.html}.}  rather than a transformer model.
    \item \textbf{transformer:} Uses a unigram tokenizer with 5000 sub-word units.
    It features an acoustic model based on a transformer architecture~\citep{Vaswani/2017/NIPS, wolf/2020/EMNLP}, which uses attention mechanisms for both encoding and decoding to weight the influence of different parts of the input data. 
    Additionally, it employs the same pre-trained language model as seq2seq~1.   
    This architecture has achieved remarkable success across various domains, including ASR, by effectively capturing the contextual relationships in the data.
\end{itemize}

The architecture and training parameters of each model are detailed in \citet{ravanelli2021speechbrain}.
For each architecture, we employed three different training regimes:
\begin{itemize}
    \item \textbf{No augmentation}: This setting serves as a control. 
    The models are trained on a clean dataset without any augmentation to estimate a baseline performance. 
    \item \textbf{Speed-augmented}: 
    This setting introduces temporal variability in the training data by applying speed perturbations that simulate natural speech tempo variations~\citep{ko2015audio}.
    \item \textbf{Fully augmented}: 
    For data augmentation in this regime, speed variations are combined with background noises and reverberations to mimic challenging and realistic acoustic environments~\citep{park2019specaugment, park2020specaugment}.
\end{itemize}

\subsection{Dataset}
We use the LibriSpeech dataset, a widely recognized corpus for ASR research that includes approximately thousand hours of read English speech from audiobooks~\citep{panayotov2015librispeech}. 
It is often used as a benchmark for evaluating speech recognition models. 
Our models are trained on the first 100 hours of the clean LibriSpeech dataset, which provides a well-balanced platform for conducting controlled, repeatable, and computationally feasible experiments on adversarial robustness across diverse ASR model types.
Note that some pre-trained models, like CTC, may have been exposed to larger portions of LibriSpeech.

\subsection{Data Augmentation}
\label{subsection:data_augmentation}
For speech augmentation, we apply several techniques to artificially corrupt the clean audio signals provided by SpeechBrain. 
These techniques include:
\begin{itemize}
    \item \textbf{Reverberation}: To $10\%$ of the files, we apply reverberation to simulate echo effects.
    \item \textbf{Background noise}: Random samples from the Freesound portion of the MUSAN corpus, which include $843$ recordings of music, speech, and background noises~\citep{ko2017study}.
    The recordings are incorporated into the clean signal at random signal-to-noise ratios.
    \item \textbf{Speed variations}: The audio signal is re-sampled at three different speeds: reduced by $5\%$, original rate, and increased by $5\%$.
\end{itemize}

\begin{figure*}[!t]
    \centering
    \begin{subfigure}[b]{0.32\textwidth}
        \centering
        \includegraphics[width=\textwidth]{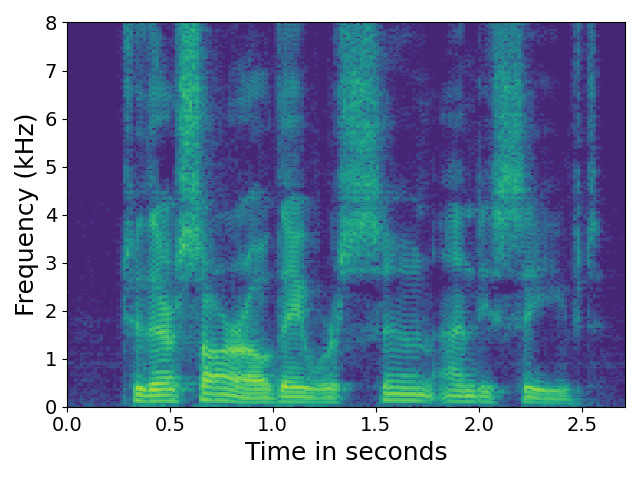}
        \caption{Original audio signal. Transcription: "to their sorrow they were soon undeceived".}
    \end{subfigure}
    \hfill
    \begin{subfigure}[b]{0.32\textwidth}
        \centering
        \includegraphics[width=\textwidth]{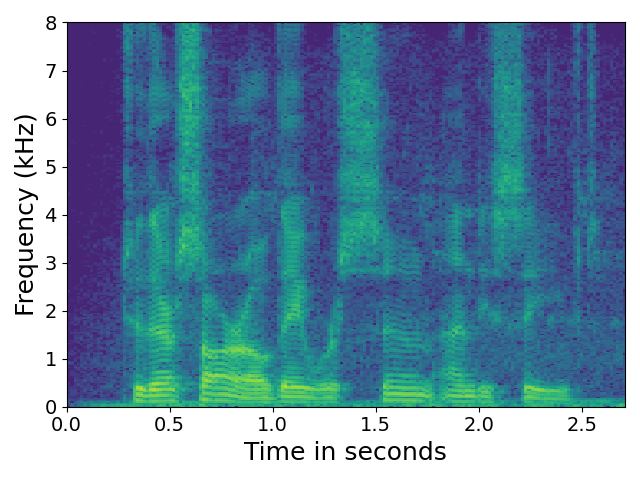}
        \caption{Noisy original audio signal. Transcription: "to their sorrow they were assumes under seats".}
    \end{subfigure}
    \hfill
    \begin{subfigure}[b]{0.32\textwidth}
        \centering
        \includegraphics[width=\textwidth]{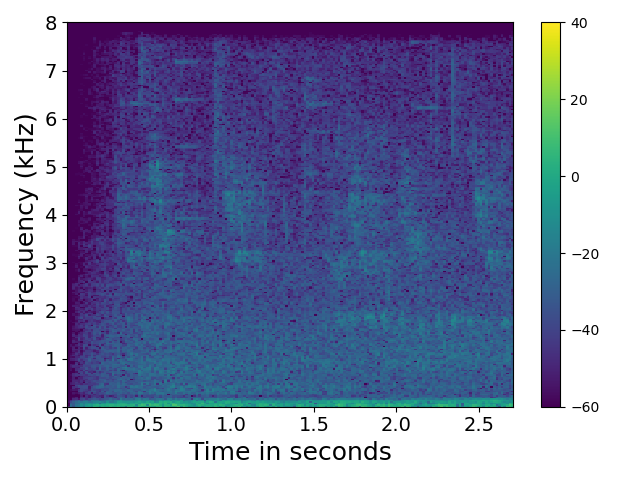}
        \caption{Noise signal.
        Difference between (a) and (b).}
        \medskip
        \smallskip
    \end{subfigure}
        \caption{The spectrogram of the original audio signal (a) is compared to the spectrogram of its corresponding noisy example, that in this case is constructed by adding background noise to the original audio signal (b) and the spectrogram of the noise (c), which is the difference between the original and its noisy counterpart.
        }
        \label{fig:benign_noisy}
\end{figure*}

Figure~\ref{fig:benign_noisy} illustrates the spectrum of a benign audio signal and its corresponding noisy counterpart. 
See \ref{app2} for the code snippet that shows how the corruptions are employed.

\subsection{Adversarial Attacks}
In our experiments, we utilize the C\&W white-box attack, a prominent adversarial technique initially designed to attack the DeepSpeech ASR system, see Section~\ref{subsec1}.
We generate 300 adversarial examples per architecture and per model.
In addition, we investigate two black-box attacks: the Alzantot and the Kenansville attack, for each of which we create 100 adversarial examples per architecture and per model. 

For C\&W and Alzantot attacks, we base our implementation on the \texttt{robust\_speech} framework~\citep{olivier2022recent}, adhering to the parameters defined therein for consistency and comparability, with the maximum number of iterations set to $3000$. 
These methodologies allow us to systematically assess the vulnerability of SpeechBrain models to sophisticated adversarial threats.
For the Kenansville attack, we use its online implementation\footnote{Kenansville attack code, see \url{https://github.com/kwarren9413/kenansville_attack}.} and run 15 iterations.

\subsection{Evaluation metrics}
To evaluate the ASR performance on benign, noise-augmented, or adversarial files, we use the following two metrics:
\begin{itemize}
    \item \textbf{Word Error Rate (WER)}: WER is a standard ASR performance metric that compares the recognized word sequence with the reference text. 
    It is defined as:
    \begin{equation}
        \text{WER} = 100 \cdot \frac{S+D+I}{N} \enspace,
    \label{eq1}
    \end{equation}
    where $S$, $D$, and $I$ represent the numbers of substituted, deleted, and inserted words, respectively, and $N$ denotes the total word count of the reference text.
    \item \textbf{Success Rate}: The success rate is the percentage of samples correctly recognized as the given reference text. 
    It is calculated as:
    \begin{equation}
        \text{Succ.\ rate} = 100 \cdot \frac{N_{S}}{N_T}\enspace,
    \label{eq2}
    \end{equation}
    where $N_S$ is the number of audio clips with no transcription error compared to the reference text, out of a total of $N_T$ examples. 
    For benign samples and black-box adversarial attacks, the ground truth serves as the reference, whereas for C\&W adversarial examples, we use the target transcription. 
    For adversarial success rate, we distinguish between two cases: for targeted attacks, the reference is the target sentence; for untargeted attacks, any transcription differing from the ground truth sentence counts as an adversarial success.
\end{itemize}

In order to quantify the distortion in adversarial audio files, we use the following metrics.
Let $x(t)$ denote the original continuous audio file and $X(k)$ denote the discretized version:
\begin{itemize}
    \item \textbf{dB\(_{\text{x}}\)}: \citet{carlini2018audio} quantify the distortion of adversarial examples by first calculating the relative loudness of an audio signal $x$ given by
    \begin{align*}
        \text{dB}(x) = 20 \cdot \max_{t} \log_{10}x(t) \enspace{.}
    \end{align*}
    Subsequently, the distortion level is determined as the difference between the relative loudness of the perturbation $\delta$ and the audio signal $x$, defined as:
    \begin{align*}
     \text{dB}_{\text{x}}(\delta) = \text{dB}(x) - \text{dB}(\delta) \enspace{,}
    \end{align*}
    assuming equal lengths of $x$ and $\delta$.
    \item \textbf{SI-SDR}: The Scale-Invariant Signal-to-Distortion Ratio improves upon the traditional non-segmental Signal-to-Noise Ratio by considering the scale-invariance of audio signals~\citep{Roux2019ICASSP}. 
    It measures, in decibels, the ratio of the energies of the reference sources to the distortion signal, while accounting for any scaling differences between the two signals.
    The SI-SDR is calculated as
    \begin{align*}
        \text{SI-SDR} = 10 \cdot \log_{10} \frac{\sum_{k} (\alpha \cdot X(k))^2}{\sum_{t} (\alpha \cdot X(k) - \hat{X}(k))^2} \enspace{,}
    \end{align*}
    where $X$ represents the benign example, and $\hat{X}$ is its adversarial counterpart and the scaling factor $\alpha$ is given by
    \begin{align*}
        \alpha =  \frac{\sum_{k} X(k) \cdot \hat{X}(k)} {\sum_{k}X(k)^2} \enspace{.}
    \end{align*}
    The calculation assumes that $X$ and $\hat X$ are of equal length.
    \item \textbf{SNR\(_\text{Seg}\)}: The Segmental Signal-to-Noise Ratio quantifies noise energy in decibels over the entire audio signal. 
    This metric is determined by calculating energy ratios of individual frames and averaging them, aligning more closely with human auditory perception compared to the non-segmental approach~\citep{mermelstein1979evaluation}. 
    The SNR\(_\text{Seg}\) is determined as:
    \begin{equation*}
        \text{SNR\(_{\text{Seg}}\)} = \frac{10}{M} \cdot  \sum_{m=0}^{M-1}\log_{10}\frac{\sum_{k=mF}^{mF+F-1}X(k)^2}{\sum_{k=mF}^{mF+F-1}\delta(k)^2}  \text{,} \enspace
    \end{equation*}
    where \(M\) represents the number of frames in a signal, \(F\) is the frame length, \(X\) denotes the clean audio signal, and \(\delta\) the adversarial perturbation, assuming equal lengths of \(X\) and \(\delta\).
\end{itemize}

\subsection{Statistical Tests}
To assess whether adversarial samples generated from two different training regimes share the same underlying distribution, we perform the Kolmogorov-Smirnov (KS) test. 
This non-parametric test compares the distributions of two samples and tests the null hypothesis that they are drawn from the same distribution. 
The hypothesis for the KS test is as follows:
\begin{itemize}
    \item \textbf{Null hypothesis}: The two distributions are identical.
    \item \textbf{Alternative hypothesis}: The two distributions are different.
\end{itemize}
A significance level of $\alpha = 0.05$ is used for the test. 
If the p-value is less than or equal to $0.05$ ($p \leq 0.05$), and the Kolmogorov-Smirnov statistic exceeds $0.1$, it indicates that the distributions are significantly different. 
On the other hand, if the p-value is greater than 0.05 ($p > 0.05$), we fail to reject the null hypothesis, implying that there is no statistically significant difference between the two distributions.
Therefore, if the KS test reveals a statistically significant difference, it suggests that the adversarial samples from the two regimes differ significantly. 
This could imply a noticeable change in model robustness due to the differences in training regimes. 
For instance, if one regime shows significantly better adversarial performance, it may indicate that its training setup contributes positively to the model's robustness.

\section{Experiments}
This section outlines the experiments conducted to evaluate the robustness of ASR models against noise and adversarial attacks.

\subsection{Evaluating Noise Robustness}
In our initial experiment, we assess the effectiveness of noise-augmented training on enhancing noise robustness in SpeechBrain models. 
For each architecture, we verify whether robustness training improves the models' performance against natural acoustic perturbations. 
For this, we add background noise to every sample according to the guidelines specified in the ``Background Noise'' from subsection~\ref{subsection:data_augmentation}. 
For detailed information on the parameter setup, refer to \ref{app:noiseAugmentation}.

In evaluating noise robustness, the reference text for WER and success rate calculations is the ground-truth transcript of the clean test sample. 
When training ARS models, we aim for models that exhibit low WER and high benign success rate on benign and noise-augmented data, indicating accurate recognition of the ground-truth transcription. 

\subsection{Evaluating Adversarial Robustness}
To assess adversarial robustness, we first generated adversarial examples for the three different attack types (C\&W, Alzantot, and Kenansville) described above.
To assess the strength of the adversarial examples, we compare the success rate of adversarial attacks against our models (see Eq.~\ref{eq2}) and analyze the WER (see Eq.~\ref{eq1}). 
For the untargeted black-box adversarial attacks, the ground truth serves as the reference; that is, the attack is considered stronger for a higher WER and lower success rate. 
For adversarial examples generated by the targeted white-box attack C\&W, we use the target transcription as ground truth. 
Thus, the lower the WER and the higher the success rate, the stronger the attack.

Additionally, for the C\&W and Alzantot attacks, we evaluate the computational complexity of generating adversarial examples by recording the first hit, defined as the number of iterations required to craft the first successful adversarial example.

We employ three distinct types of distortion measurements to quantify the imperceptibility of adversarial perturbations.
If the perturbations are perceptible by humans, they can easily be spotted.
Higher values of dB\(_{\text{x}}\), SI-SDR, and SNR\(_\text{Seg}\) indicate a lower amount of added noise. 
In the context of evaluating adversarial robustness in ASR systems, we expect these values to be lower for models trained with augmentation methods, indicating that these models require more noise for adversarial samples to succeed.

Tracking this set of different metrics allows us to dissect the nuances of adversarial robustness, understanding the trade-offs between attack efficacy, computational expense, and perceptibility in the context of our SpeechBrain models.

\section{Results}
In the following, we present and discuss the results of our experiments. A demo is available at \url{https://matiuste.github.io/AssesNoiseAugm/}, where we also provide pre-trained models to support reproducibility.
These models can be used to generate adversarial examples and evaluate the effectiveness of noise augmentation methods.

\begin{table*}[t!]
    \centering    
    \footnotesize
    \begin{tabular}{c c : r r : r r : r}
    \multicolumn{2}{c}{} & 
    \multicolumn{2}{:c:}{\textbf{Benign Data}} & 
    \multicolumn{2}{c:}{\textbf{Noisy Data}} &
    \multicolumn{1}{c}{\textbf{}} \\
        \textbf{Model} & \textbf{Augm.} &  \textbf{Succ.\ rate {\color{gray}$\uparrow$}} & \textbf{WER {\color{gray}$\downarrow$}} & \textbf{Succ.\ rate {\color{gray}$\uparrow$}} & \textbf{WER {\color{gray}$\downarrow$}} & \textbf{$\Delta$ WER} \\
    \hline
        \multirow{3}{*}{CTC} & No augmentation & 85.67\% & 3.11\% & 81.33\% & 5.39\% &  2.28\% \\
        & Speed-augmented & 87.00\% & 2.84\% & \textbf{83.00\%} & \textbf{5.06\%} & 2.22\% \\
        & Fully augmented & 86.33\% & 2.95\% & 82.33\% & 5.12\% & 2.17\% \\
    \hdashline
        \multirow{3}{*}{Seq2seq 1} & No augmentation & 70.33\% & 8.18\% & 54.33\% & 19.52\% & 11.35\% \\
        & Speed-augmented & 71.33\% & 8.06\% & 55.67\% & 20.02\% & 11.96\% \\
        & Fully augmented & 71.00\% & 8.57\% & \textbf{65.33\%} & \textbf{11.51\%} & 2.95\% \\
    \hdashline
        \multirow{3}{*}{Seq2seq 2} & No augmentation & 67.00\% & 9.23\% & 53.33\% & 21.08\% & 11.85\% \\
        & Speed-augmented & 71.33\% & 8.18\% & 54.67\% & 20.02\% & 11.85\% \\
        & Fully augmented & 66.67\% & 10.01\% & \textbf{56.67\%} & \textbf{14.35\%} & 4.34\% \\
    \hdashline
        \multirow{3}{*}{Transformer} & No augmentation & 65.00\% & 9.90\% & 52.33\% & 19.91\% & 10.01\% \\
        & Speed-augmented & 66.67\% & 9.29\% & 53.00\% & 19.47\% & 10.18\% \\
        & Fully augmented & 68.00\% & 8.51\% & \textbf{53.67\%} & \textbf{18.85\%} & 10.34\% \\
    \end{tabular}
    \caption{Performance of the considered ASR systems on benign and noisy data, averaged over 300 utterances. 
    WER and success rate are measured w.r.t.\ the ground truth transcription.
    The arrows indicate the direction for models that are more noise robust, e.g., models with a higher success rate are considered more robust against noisy examples.
    The bold numbers indicate the most robust system per model with respect to the chosen parameter.
    }
    \label{tab:BenignResults}
\end{table*}

\subsection{Evaluating Noise Robustness}
To evaluate noise robustness, WER and success rate are calculated using the ground-truth transcript of the clean test sample. 
Low WER and high success rate indicate accurate transcription, allowing us to assess the effectiveness of noise-augmented training before analyzing its impact on adversarial robustness.

Our experimental results, see Table~\ref{tab:BenignResults}, corroborate that models subjected to noise-augmented training exhibit enhanced robustness when confronted with speech samples with environmental noise. 
For each model considered, the success rate and WER for noisy data have improved with noise-augmented training (the most robust models are highlighted in bold). 
This is evidenced by a lower WER in comparison to their counterparts trained without any augmentation.

This improvement is particularly pronounced for the seq2seq models, which build on minimal pre-trained components, while the enhancement is less apparent for the transformer-based models (including CTC) that rely more on pre-trained parts, especially for feature extraction. 
Interestingly, the CTC-based models maintain a high success rate even with noisy data, whereas for all other architectures, the success rate drops significantly. 
This could be attributed to the fact that the CTC models incorporate pre-trained models that were exposed to noisy data during training. 
These findings indicate that noise-augmented training indeed enhances noise robustness, although the extent of improvement varies depending on the architecture.

\begin{table*}[t!]
    \centering    
    \footnotesize
    \begin{tabular}{c c : r r r r r r}
    \multicolumn{2}{c}{} & 
    \multicolumn{6}{:c}{\textbf{C\&W Adversarial Data}} \\
        \textbf{Model} & \textbf{Augm.} & \textbf{Succ.\ rate {\color{gray}$\downarrow$}} & \textbf{WER {\color{gray}$\uparrow$}} & \textbf{First hit {\color{gray}$\uparrow$}} & \textbf{$\text{dB\(_{\text{x}}\)}$ {\color{gray}$\downarrow$}} & \textbf{SI-SDR {\color{gray}$\downarrow$}} & \textbf{$\text{SNR}_{\text{seg}}$ {\color{gray}$\downarrow$}} \\
    \hline
        \multirow{3}{*}{CTC} & No augmentation & 100.0\% & 0.00 & 155 &  35.41 & 29.29 & 18.54 \\
        & Speed-augmented & \textbf{99.6\%} & \textbf{0.05} & \underline{\textbf{343}} & \underline{\textbf{29.59}} & \underline{\textbf{25.30}} & \underline{\textbf{14.08}} \\
        & Fully augmented & 100.0\% & 0.00 & 158 & 35.40 & 29.04 & 18.28  \\
    \hdashline
        \multirow{3}{*}{Seq2seq 1} & No augmentation & 88.3\% & 5.53 & 454 & 27.21 & 20.29 & 9.34 \\
        & Speed-augmented & 86.0\% & 7.16 & 323 & 27.46 & 20.42 & 9.47 \\
        & Fully augmented & \textbf{74.0\%} & \underline{\textbf{14.89}} & \textbf{479} & \underline{\textbf{23.92}} & \textbf{\underline{17.74}} & \underline{\textbf{6.69}}  \\
    \hdashline
        \multirow{3}{*}{Seq2seq 2} & No augmentation & 98.3\% & 0.47 & 107 & 41.99 & 30.23 & 19.37 \\
        & Speed-augmented & 99.6\% & 0.05 & 113 & 41.13 & 29.65 & 18.84 \\
        & Fully augmented & \textbf{96.0\%} & \textbf{1.53} & \underline{\textbf{186}} & \textbf{\underline{37.65}} & \textbf{\underline{26.88}} & \textbf{\underline{15.93}} \\
    \hdashline
        \multirow{3}{*}{Transformer} & No augmentation & 100.0\% & 0.00 & 19 & 56.26 & 40.77 & 28.89 \\
        & Speed-augmented & 100.0\% & 0.00 & 18 & \underline{\textbf{54.52}} & \textbf{\underline{39.10}} & \underline{\textbf{27.23}} \\
        & Fully augmented & 100.0\% & 0.00 & \underline{\textbf{20}} & \underline{54.90} & \underline{39.47} & \underline{27.59}  \\
    \end{tabular}
    \caption{Results for C\&W targeted attack. 
    All values are averaged over 300 adversarial examples. 
    WER and success rate are measured w.r.t.\ the target adversarial transcription.
    The arrows indicate the direction for models that are more robust against adversarial examples, e.g., models with a lower success rate are considered more robust against adversarial examples.
    The bold numbers indicate the more robust model with respect to the chosen parameter.
    Underlined numbers indicate a statistically significant difference based on the KS test between the no augmentation training regime and the evaluated augmented training regime.
    }
    \label{tab:CWResults}
\end{table*}

\subsection{Evaluating Adversarial Robustness}
Table~\ref{tab:CWResults} encapsulates the success rate, computational effort, and perceptibility metrics across various architectures and models for the targeted C\&W~attack. 
In this setting, WER and success rate are computed using the target adversarial transcript, where higher WER and lower success rate indicate stronger adversarial robustness.
For distortion metrics, lower values of $\text{dB}_{x}$, $\text{SI-SDR}$, and $\text{SNR}_{\text{seg}}$ suggest that greater noise is needed for the adversarial samples to succeed, reflecting better model resistance. 

The results reveal that for the seq2seq models trained from scratch, those that underwent noise-augmented training demonstrate a slightly reduced adversarial success rate, a higher WER, or more perceptible noise, indicating improved robustness to adversarial attacks. 
As shown in Table~\ref{tab:CWResults}, the underlined values indicate cases where distortion metrics differ statistically significantly between the baseline (no augmentation) and data augmentation regimes, indicating that data augmentation leads to noisier adversarial examples.
While improvements in WER and success rate are modest, all augmentation regimes consistently enhance adversarial robustness by forcing adversarial examples to be noisier. 
This effect is most pronounced in seq2seq-based models, which benefit significantly from augmentation. 
Transformer-based models (including the CTC model) show more limited gains but still exhibit increased resistance to adversarial attacks, mirroring the trends observed in noise robustness.
Interestingly, we did not find a clear connection between the model's WER on the benign data (see Table~\ref{tab:BenignResults}, fourth column) and the adversarial success rate (see Table~\ref{tab:CWResults}, third column).

\begin{table*}[t!]
    \centering
    \footnotesize
    \begin{tabular}{c c : r r r r r}
    \multicolumn{2}{c}{} & 
    \multicolumn{5}{c}{\textbf{Alzantot Adversarial Data}} \\
    \textbf{Model} & \textbf{Augm.} & 
    \textbf{Succ.\ rate {\color{gray}$\downarrow$}} & \textbf{WER {\color{gray}$\downarrow$}} & \textbf{First hit {\color{gray}$\uparrow$}} & \textbf{SI-SDR {\color{gray}$\downarrow$}} & \textbf{$\text{SNR}_{\text{seg}}$ {\color{gray}$\downarrow$}} \\
    \hline
        \multirow{3}{*}{CTC} &  No augmentation & \textbf{58.00\%} & \textbf{10.63\%} & 77 & 20.75 & 12.25 \\
        & Speed-augmented & 61.00\% & 12.18\% & \textbf{\underline{91}} & \textbf{18.95} & \textbf{8.13} \\ 
        & Fully augmented & 60.00\% & 10.94\% & 56 & 21.28 & 12.10 \\
    \hdashline
        \multirow{3}{*}{Seq2seq 1} & No augmentation & 96.00\% & 45.92\% & 42 & \textbf{18.61} & \textbf{8.15} \\ 
        & Speed-augmented & 95.00\% & 42.72\% & 37 & 19.40 & 8.72 \\ 
        & Fully augmented & \textbf{86.00\%} & \textbf{\underline{29.31\%}} & \textbf{77} & 18.97 & 8.52 \\ 
    \hdashline
        \multirow{3}{*}{Seq2seq 2} & No augmentation & 96.00\% & 51.29\% & 37 & \textbf{17.86} & \textbf{7.05} \\ 
        & Speed-augmented & 99.00\% & 47.78\% & 37 & 18.66 & 8.05 \\ 
        & Fully augmented & \textbf{91.00\%} & \textbf{\underline{33.13\%}} & \textbf{\underline{72}} & 18.34 & 7.94 \\ 
    \hdashline
        \multirow{3}{*}{Transformer} & No augmentation & 99.00\% & 39.63\% & 42 & \textbf{19.99} & \textbf{9.24} \\ 
        & Speed-augmented & \textbf{96.00\%} & 38.18\% & \textbf{50} & 20.80 & 11.24 \\ 
        & Fully augmented & 99.00\% & \textbf{36.74\%} & 34 & 20.86 & 11.12 \\ 
    \end{tabular}
    \caption{Results for the Alzantot untargeted attack.
    Values are averaged over 100 adversarial examples.
    For this untargeted attack, when calculating the adversarial success rate, any transcription but the ground truth transcription is considered a success.
    For simplicity, the WER is measured w.r.t.\ the ground truth transcription. 
    It thus provides an insight of how different the resulting adversarial samples are from the ground truth.
    Again, the arrows indicate the direction for models that are more robust against adversarial examples.
    The bold numbers indicate the more robust model with respect to the chosen parameter.
    Underlined numbers indicate a statistically significant difference based on the KS test between the no augmentation training regime and the evaluated augmented training regime.}
    \label{tab:Alzantot}
\end{table*}
\begin{table*}[t!]
    \centering
    \footnotesize
    \begin{tabular}{c c : r r r r r}
    \multicolumn{2}{c}{} & 
    \multicolumn{5}{c}{\textbf{Kenansville Adversarial Data}} \\
    \textbf{Model} & \textbf{Augm.} & 
    \textbf{Succ.\ rate {\color{gray}$\downarrow$}} & \textbf{WER {\color{gray}$\downarrow$}} & \textbf{dB\(_{\text{x}}\) {\color{gray}$\downarrow$}} & \textbf{SI-SDR {\color{gray}$\downarrow$}} & \textbf{$\text{SNR}_{\text{seg}}$ {\color{gray}$\downarrow$}}\\
    \hline
        \multirow{3}{*}{CTC} &  No augmentation & 80.00\% & \textbf{13.62\%} & 12.90 & 14.94 & 6.58 \\
        & Speed-augmented & 99.00\% & 17.75\% & \textbf{10.13} &  \textbf{12.58} & \textbf{4.56} \\ 
        & Fully augmented & \textbf{78.00\%} & 13.83\% & 11.89 & 14.32 & 5.98 \\
    \hdashline
        \multirow{3}{*}{Seq2seq 1} & No augmentation & \textbf{92.00\%} & \textbf{21.88\%} & 19.28 & 21.39 & 12.16 \\ 
        & Speed-augmented & 96.00\% & 24.97\% & 18.69 & 20.72 & 11.46 \\ 
        & Fully augmented & 99.00\% & 22.81\% & \textbf{\underline{17.21}} & \textbf{\underline{19.44}} & \textbf{10.56} \\ 
    \hdashline
        \multirow{3}{*}{Seq2seq 2} & No augmentation & 97.00\% & 31.37\% & 18.38 & 20.80 & 11.67 \\ 
        & Speed-augmented & \textbf{94.00\%} & \textbf{24.77\%} & 19.11 & 21.49 & 12.32 \\ 
        & Fully augmented & 96.00\% & 26.83\% & \textbf{\underline{15.83}} & \textbf{18.34} & \textbf{\underline{9.56}} \\ 
    \hdashline
        \multirow{3}{*}{Transformer} & No augmentation & \textbf{96.00\%} & 24.46\% & 20.75 & 22.77 & 13.46 \\ 
        & Speed-augmented & 97.00\% & 22.81\% & \textbf{19.56} &  \textbf{21.80} & \textbf{12.60} \\ 
        & Fully augmented & \textbf{96.00\%} & \textbf{21.57\%} & 20.22 & 22.32 & 13.02 \\ 
    \end{tabular}
    \caption{Results for the Kenansville untargeted attack.
    All values are averaged over 100 adversarial examples.
    For this untargeted attack, when calculating the adversarial success rate, any transcription but the ground truth transcription is considered a success.
    For simplicity, the WER is measured w.r.t.\ the ground truth transcription. 
    It thus provides an insight of how different the resulting adversarial samples are from the ground truth.
    The arrows indicate the direction for models that are more robust against adversarial examples.
    The bold numbers indicate the more robust model with respect to the chosen parameter.
    Underlined numbers indicate a statistically significant difference based on the KS test between the no augmentation training regime and the evaluated augmented training regime.}
    \label{tab:Kenansville}
\end{table*}
While from a privacy and security perspective, the most relevant attacks are targeted, we also explored the impact of noise-augmented training on the model robustness w.r.t.~untargeted black-box attacks.
In this context, WER and success rate are calculated using the ground-truth transcript, where lower WER and success rate reflect stronger adversarial robustness.
While the success rate defines the attacks' success, the WER is a general measure of transcription accuracy.
For distortion metrics, lower values indicate that adversarial samples need more noise to succeed, demonstrating better model resilience.
Here, the benefits of noise-augmented training are less definitive, as detailed in Tables~\ref{tab:Alzantot}~and~\ref{tab:Kenansville}. 
For the Alzantot attack, models trained with augmented data exhibit commendable performance in terms of success rate and WER, with the seq2seq augmented models displaying significant improvements, as highlighted by the underlined WER values in Table~\ref{tab:Alzantot}. 
These models demonstrate enhanced robustness, as confirmed by the Kolmogorov-Smirnov test, showing statistically significant differences compared to the no augmentation counterparts. 
While models trained without augmentation show a slight advantage in distortion metrics like segmental SNR, this difference is not statistically significant.
For the Kenansville attack, differences in success rate and WER between augmented and non-augmented models are not statistically significant. However, models trained with augmented data consistently show lower distortion metrics, indicating improved robustness. 
The seq2seq models, in particular, demonstrate statistically significant improvements, as underlined in Table~\ref{tab:Kenansville}. 
Overall, augmented models are more resistant to adversarial attacks, requiring adversarial examples to be noisier in order to succeed.

In summary, our findings indicate that incorporating speed or noise augmentation into the training process enhances the adversarial robustness of most models when confronted with targeted white-box attacks as well as untargeted black-box attacks. 
In cases where data augmentation does not lead to higher robustness compared to the base model, it also does not result in a severe robustness reduction. 
This underlines that noise augmentation should be used as a standard routine in the training of neural network-based ASR systems.

\section{Discussion and Future Work}
Our findings reveal that across the board, ASR models trained with noise augmentation -- encompassing background noise, speed variations, and reverberations -- exhibit enhanced robustness against adversarial attacks compared to those without such augmentation. 
Interestingly, the implementation of speed variations alone as a form of augmentation also contributed positively to the robustness of the models, albeit to a lesser extent than the combination of multiple noise types. 
This hints that robustness gains may be attributed to the increased diversity and volume of the dataset.
Through our investigation, it becomes evident that incorporating speed and noise augmentation into the training pipeline is beneficial for both noise robustness and adversarial robustness.

The findings from our study suggest several intriguing directions for further research. 
First, while noise augmentation has proven effective, understanding the optimal combination and intensity of different augmentation types remains an open question. 
The relative contributions of background noise, speed variations, and reverberations need to be quantified to develop more targeted augmentation strategies.
Additionally, the role of dataset size and diversity in enhancing robustness merits deeper exploration. 
Our results indicate that simply increasing the dataset size with varied augmentations can lead to significant robustness improvements. 
Future studies could systematically vary the size and diversity of training data to better understand this relationship.
In particular, this should encompass varying levels of signal-to-noise ratio and different noise dataset sizes.
Moreover, our research highlights the complex interplay between different forms of augmentation and their impact on adversarial robustness and noise robustness.
In particular, it could be relevant to study what impact noise-augmented training has on clean speech quality or which artifacts may be introduced by noise-augmented training (and potentially exploited by adversarial attacks).
A thorough investigation into how these augmentations interact with the robustness w.r.t.~various attack types could provide more granular insights into defense mechanisms.

For future work, we believe that exploring the interplay between noise augmentation and other defense strategies, such as adversarial training and model ensembling, could yield a more robust and multi-faceted approach to securing ASR systems against adversarial attacks. 
By continuing to refine augmentation techniques and exploring their integration with other defensive measures, we can move towards developing ASR systems that are both robust and reliable in the face of sophisticated adversarial challenges.

\section{Conclusion}
In conclusion, our study provides a foundational understanding of the benefits of noise augmentation for ASR systems. 
Our comprehensive study has shed light on the efficacy of noise augmentation as a strategy for improving the robustness of automatic speech recognition systems with respect to noise as well as to adversarial examples.
By conducting a systematic analysis of four ASR architectures, each trained with different augmentation techniques and subjected to a spectrum of adversarial 
attacks, we have provided empirical evidence supporting the benefits of noise augmentation for adversarial robustness.

\section{Acknowledgements}
This work was partly supported by the Deutsche Forschungsgemeinschaft (DFG, German Research Foundation) under Germany's Excellence Strategy -- EXC 2092 CASA -- 390781972 and the Wilhelm and Günter Esser Foundation.

\appendix

\section{Noise Augmentation in Training}
\label{app2}
The following code snippet displays the corruptions used for the non-augmented model.
\begin{lstlisting}[style=yaml]
env_corrupt: !new:speechbrain.lobes.augment.EnvCorrupt
  openrir_folder: <folder>
  babble_prob: 0.0
  reverb_prob: 0.0
  noise_prob: 0.0
  noise_snr_low: 0
  noise_snr_high: 15
  
augmentation: !new:speechbrain.lobes.augment.TimeDomainSpecAugment
  sample_rate: 16000
  speeds: [100]
\end{lstlisting}

The following code snippet displays the corruptions used for the speed-augmented model.
\begin{lstlisting}[style=yaml]
env_corrupt: !new:speechbrain.lobes.augment.EnvCorrupt
  openrir_folder: <folder>
  babble_prob: 0.0
  reverb_prob: 0.0
  noise_prob: 0.0
  noise_snr_low: 0
  noise_snr_high: 15
  
augmentation: !new:speechbrain.lobes.augment.TimeDomainSpecAugment
  sample_rate: 16000
  speeds: [95, 100, 105]
\end{lstlisting}

The following code snippet displays the corruptions used for the fully augmented model.
\begin{lstlisting}[style=yaml]
env_corrupt: !new:speechbrain.lobes.augment.EnvCorrupt
  openrir_folder: <folder>
  babble_prob: 0.1
  reverb_prob: 0.1
  noise_prob: 1.0
  noise_snr_low: 0
  noise_snr_high: 15
  
augmentation: !new:speechbrain.lobes.augment.TimeDomainSpecAugment
  sample_rate: 16000
  speeds: [95, 100, 105]
\end{lstlisting}

\section{Noise Augmentation in Testing}\label{app:noiseAugmentation}
The following code snippet displays the corruptions used in the augmentation for testing robustness.
\begin{lstlisting}[style=yaml]
env_corrupt: !new:speechbrain.lobes.augment.EnvCorrupt
  openrir_folder: <folder>
  babble_prob: 0.0
  reverb_prob: 0.0
  noise_prob: 1.0
  noise_snr_low: 0
  noise_snr_high: 15
augmentation: !new:speechbrain.lobes.augment.TimeDomainSpecAugment
  sample_rate: 16000
  speeds: [100]
\end{lstlisting}

\bibliographystyle{elsarticle-harv} 
\bibliography{CSL_Journal}





\end{document}